\documentstyle[12pt]{article}

\newcommand{\sect}[1]{\setcounter{equation}{0}\section{#1}}
\renewcommand{\theequation}{\thesection.\arabic{equation}}

\def\bseq{\begin{subequation}}  
\def\eseq{\end{subequation}}
\def\bsea{\begin{subeqnarray}}  
\def\esea{\end{subeqnarray}}

\def\beq{\begin{equation}}
\def\eeq{\end{equation}}
\def\eea{\end{eqnarray}}
\def\bq{\begin{quote}}
\def\eq{\end{quote}}

\newcommand{\EQ}{\begin{equation}}
\newcommand{\EN}{\end{equation}}
\newcommand{\bea}{\begin{eqnarray}}
\newcommand{\ena}{\end{eqnarray}}

\renewcommand{\a}{\alpha}

\newcommand{\th}{\theta}

\newcommand{\pa}{\partial}
\newcommand{\g}{\gamma}

\renewcommand{\l}{\lambda}
\renewcommand{\L}{\Lambda}
\newcommand{\m}{\mu}

\newcommand{\n}{\nu}

\newcommand{\s}{\sigma}

\def\Mb{\kern 2pt\mathchoice
            {
                 \vbox{\hrule width10pt height 0.4pt depth 0pt
                 \kern 1.2pt\hbox{\kern -2pt$\displaystyle M$}}}
            {
                 \vbox{\hrule width10pt height 0.4pt depth 0pt
                 \kern 1.2pt\hbox{\kern -2pt$\textstyle M$}}}
            {
                 \vbox{\hrule width6pt height 0.4pt depth 0pt
                 \kern 1.0pt\hbox{\kern -2pt$\scriptstyle M$}}}
            {
                 \vbox{\hrule width5pt height 0.4pt depth 0pt
                 \kern 0.8pt\hbox{\kern -2pt$\scriptscriptstyle M$}}}}

\def\Sb{\kern 2pt\mathchoice
            {
                 \vbox{\hrule width6pt height 0.4pt depth 0pt
                 \kern 1.2pt\hbox{\kern -2pt$\displaystyle S$}}}
            {
                 \vbox{\hrule width6pt height 0.4pt depth 0pt
                 \kern 1.2pt\hbox{\kern -2pt$\textstyle S$}}}
            {
                 \vbox{\hrule width3.5pt height 0.4pt depth 0pt
                 \kern 1.0pt\hbox{\kern -2pt$\scriptstyle S$}}}
            {
                 \vbox{\hrule width3pt height 0.4pt depth 0pt
                 \kern 0.8pt\hbox{\kern -2pt$\scriptscriptstyle S$}}}}

\def\Rb{\kern 2pt\mathchoice
            {
                 \vbox{\hrule width5.5pt height 0.4pt depth 0pt
                 \kern 1.2pt\hbox{\kern -2.5pt$\displaystyle R$}}}
            {
                 \vbox{\hrule width5.5pt height 0.4pt depth 0pt
                 \kern 1.2pt\hbox{\kern -2.5pt$\textstyle R$}}}
            {
                 \vbox{\hrule width3.5pt height 0.4pt depth 0pt
                 \kern 1.0pt\hbox{\kern -2.2pt$\scriptstyle R$}}}
            {
                 \vbox{\hrule width3pt height 0.4pt depth 0pt
                 \kern 0.8pt\hbox{\kern -2.2pt$\scriptscriptstyle R$}}}}

  \def\pp{{\mathchoice
          {
              \kern 1pt%
              \raise 1pt
              \vbox{\hrule width5pt height0.4pt depth0pt
                    \kern -2pt
                    \hbox{\kern 2.3pt
                          \vrule width0.4pt height6pt depth0pt
                          }
                    \kern -2pt
                    \hrule width5pt height0.4pt depth0pt}%
                    \kern 1pt
           }
            {
              \kern 1pt%
              \raise 1pt
              \vbox{\hrule width4.3pt height0.4pt depth0pt
                    \kern -1.8pt
                    \hbox{\kern 1.95pt
                          \vrule width0.4pt height5.4pt depth0pt
                          }
                    \kern -1.8pt
                    \hrule width4.3pt height0.4pt depth0pt}%
                    \kern 1pt
            }
            {
              \kern 0.5pt%
              \raise 1pt
              \vbox{\hrule width4.0pt height0.3pt depth0pt
                    \kern -1.9pt  
                    \hbox{\kern 1.85pt
                          \vrule width0.3pt height5.7pt depth0pt
                          }
                    \kern -1.9pt
                    \hrule width4.0pt height0.3pt depth0pt}%
                    \kern 0.5pt
            }
            {
              \kern 0.5pt%
              \raise 1pt
              \vbox{\hrule width3.6pt height0.3pt depth0pt
                    \kern -1.5pt
                    \hbox{\kern 1.65pt
                          \vrule width0.3pt height4.5pt depth0pt
                          }
                    \kern -1.5pt
                    \hrule width3.6pt height0.3pt depth0pt}%
                    \kern 0.5pt
            }
        }}

  \def\mm{{\mathchoice
                       {
                             \kern 1pt
               \raise 1pt    \vbox{\hrule width5pt height0.4pt depth0pt
                                  \kern 2pt
                                  \hrule width5pt height0.4pt depth0pt}
                             \kern 1pt}
                       {
                            \kern 1pt
               \raise 1pt \vbox{\hrule width4.3pt height0.4pt depth0pt
                                  \kern 1.8pt
                                  \hrule width4.3pt height0.4pt depth0pt}
                             \kern 1pt}
                       {
                            \kern 0.5pt
               \raise 1pt
                            \vbox{\hrule width4.0pt height0.3pt depth0pt
                                  \kern 1.9pt
                                  \hrule width4.0pt height0.3pt depth0pt}
                            \kern 1pt}
                       {
                           \kern 0.5pt
             \raise 1pt  \vbox{\hrule width3.6pt height0.3pt depth0pt
                                  \kern 1.5pt
                                  \hrule width3.6pt height0.3pt depth0pt}
                           \kern 0.5pt}
                       }}

\def\pd{{\kern0.5pt
                   + \kern-5.05pt \raise5.8pt\hbox{$\textstyle.$}\kern 0.5pt}}

\def\pmd{{\kern0.5pt
                  \pm \kern-5.05pt \raise6.3pt\hbox{$\textstyle.$}\kern1.5pt}}

\def\md{\mathchoice
   {
      {{\kern 1pt - \kern-6.2pt \raise5pt\hbox{$\textstyle.$}\kern 1pt}}}
    {
      {{\kern 1pt - \kern-6.2pt \raise5pt\hbox{$\textstyle.$}\kern 1pt}}}
    {
      {\kern0.5pt - \kern-5.05pt \raise3.4pt\hbox{$\textstyle.$}\kern0.5pt}}
    {
      {\kern0.5pt - \kern-5.05pt \raise3.4pt\hbox{$\textstyle.$}\kern0.5pt}}}

\def\bone{{1}}

\def\downprop{\rule[-15pt]{0pt}{0pt}}

\newcommand{\Eh}{\hat{E}}
\newcommand{\Ec}{\check{E}}

\newcommand{\Apm}{A_+^{~-}}
\newcommand{\Amp}{A_-^{~+}}
\newcommand{\Apmd}{A_{\pd}^{~\md}}
\newcommand{\Ampd}{A_{\md}^{~\pd}}

\newcommand{\Ch}{\hat{C}}
\newcommand{\Cc}{\check{C}}
\newcommand{\Cmpp}{\Ch_{-\pd}^{~~~\pp}}
\newcommand{\Cpmp}{\Ch_{+\md}^{~~~\pp}}
\newcommand{\Cpmm}{\Ch_{+\md}^{~~~\mm}}
\newcommand{\Cmpm}{\Ch_{-\pd}^{~~~\mm}}

\newcommand{\BAE}{ \buildrel \leftarrow \over \Ec}
\newcommand{\BAH}{\buildrel \leftarrow \over H}
\newcommand{\BAEh}{ \buildrel \leftarrow \over \Eh}
\newcommand{\BAD}{ \buildrel \leftarrow \over D}

\newcommand{\Del}{\nabla}

\renewcommand{\thefootnote}{\fnsymbol{footnote}}

\begin{document}

\newpage
\begin{titlepage}
\begin{flushright}
{BRX-TH-360}
\end{flushright}
\vspace{2cm}
\begin{center}
{\bf {\large PREPOTENTIALS FOR (2,2) SUPERGRAVITY}}\\
\vspace{1.5cm}
Marcus T. Grisaru\footnote{
Work partially supported by the National Science Foundation under
grant PHY-92-22318.} \\
and\\
\vspace{1mm}
Marcia E. Wehlau\footnote{\hbox to \hsize{Current address:
Mars Scientific Consulting, 28 Limeridge Dr.,
 Kingston, ON CANADA K7K~6M3}}\\

\vspace{1mm}
{\em Physics Department, Brandeis University, Waltham, MA 02254, USA}\\
\vspace{1.1cm}
{{ABSTRACT}}
\end{center}

\bq
We present a complete solution of the constraints for two-dimensional, N=2
supergravity
in N=2 superspace. We obtain explicit expressions  for the covariant
derivatives in terms of  the vector superfield $H^m$ and,  for the two versions
of minimal (2,2) supergravity,
 a chiral   or twisted chiral scalar superfield $\phi$.
\eq

\vfill

\begin{flushleft}
September 1994

\end{flushleft}
\end{titlepage}

\newpage

\renewcommand{\thefootnote}{\arabic{footnote}}
\setcounter{footnote}{0}
\newpage
\pagenumbering{arabic}

\sect{Introduction}

In  1978 Siegel$^1$  presented the  solution of the Wess-Zumino
constraints$^2$
for four-dimensional  $N=1$ superspace supergravity. Ever since, for any
supergravity theory, one of the main endeavours of superspace practitioners has
been to solve constraints  on superspace covariant derivatives in terms of
unconstrained (pre)potentials. Solutions generally exist whenever a complete
(component) set of auxiliary fields exists.

In two dimensions, where a number of  $(p,q)$ supergravities can be
constructed, the solution to the
constraints is fairly straightforward for $(1,0)$$^3$, $(1,1)$$^4$,
and $(p,0)$$^5$ supergravity. The situation for $(2,2)$
i.e. $N=2$ supergravity is more complicated.  In principle the solution can be
obtained by direct dimensional reduction from four-dimensional $N=1$
supergravity as described in {\em Superspace}$^6$, pp. 469-472
but in practice this has not been carried out.  The solution can be obtained
easily in conformal gauge$^7$ or in light-cone gauge, but a general,
fully covariant answer has not been worked out. To the best of our knowledge
the only published attempt is by Alnowaiser$^8$. However, his work is
incomplete and some of its aspects are questionable.

We have reanalysed the problem of  obtaining prepotentials  for (2,2)
supergravity in $N=2$ superspace.  We present a complete solution  in a form
which is suitable for some applications. We achieve this, in part, by simply
working in a spinor,
light-cone basis, rather than using $\g$-matrices. Trivial as this may seem, it
allows for a more transparent treatment. In particular, it becomes obvious that
at a certain stage one has to solve a quadratic equation rather than the
quartic equation that appears in ref.~8; complete, relatively simple
results follow. We obtain explicit forms for the covariant derivatives and the
vielbein
superdeterminant.

Besides contributing to the program of finding prepotentials for all
two-dimensional supergravities, there are some additional advantages to a fully
covariant solution of the constraints. First, it may allow the study of some
situations with nontrivial (super-)Riemann surface topology where light-cone
gauge is not accessible and conformal gauge may not be
convenient. Second, it may facilitate the understanding of some issues in
induced $(2,2)$ supergravity and shed some light on its higher-loop quantum
properties. For this purpose
the development of a fully covariant background-field formalism is  desirable,
and possible$^9$. Finally, we hope that by being slightly
pedagogical, this work provides a useful review of techniques used in four
dimensions$^6$.

It is known that minimal $(2,2)$ supergravity comes in two versions, depending
whether one gauges, in addition to the Lorentz symmetry, an axial $U(1)$ or a
vector $U(1)$ tangent space symmetry$^{10,11}$. We will point out
however that the solution of the constraints for one of the theories can be
obtained easily from the solution for the other, and we will concentrate on the
axial version, which is related to four-dimensional minimal  ($n=-\frac{1}{3}$)
supergravity by dimensional reduction$^6$.

We use the following notation. Flat N=2 superspace is described by  bosonic
coordinates $x^{\pp}$ and $x^\mm$, and fermionic coordinates $\th^+$, $\th^-$,
and their complex conjugates $\th^{\pd}$ and $\th^{\md}$. The spinorial
derivatives satisfy the anticommutation relations
\EQ
\{D_+, D_{\pd}\}=i\pa_{\pp}    ~~~,~~~\{D_-,D_{\md}\}= i \pa_\mm
\EN
with all others equal to zero. We define Lorentz, $U_V(1)$ and  $U_A(1)$
tangent
space generators $\L$, $\tilde{\L}$ and $N$ by their action on spinors:
\bea
[\L, \psi_{\pm}] = \pm \frac{1}{2} \psi_{\pm} ~~~~&,&~~~~[\L, \psi_{\pmd}]
= \pm \frac{1}{2} \psi_{\pmd} \nonumber\\
{[}\tilde{\L} , \psi_{\pm}]= \mp \frac{i}{2} \psi_{\pm}~~~~&,&~~~~ [\tilde{\L},
\psi_{\pmd}]=\pm \frac{i}{2} \psi_{\pmd} \nonumber\\
{[}N , \psi_{\pm}]=  -\frac{i}{2} \psi_{\pm} ~~~~&,&~~~~ [N,
\psi_{\pmd}]=+ \frac{i}{2} \psi_{\pmd}    \ .
\ena
It will prove convenient however to define combinations of operators which act
only on undotted or dotted variables:
\EQ
M=\frac{1}{2} (\L +i \tilde{\L}) ~~~~,~~~~\Mb=\frac{1}{2} (\L -i
\tilde{\L})  \ .
\EN
Then
\bea
[M, \psi_{\pm}] = \pm \frac{1}{2} \psi_{\pm} ~~~~&,&~~~~[M, \psi_{\pmd}]=0
\nonumber\\
{[}\Mb, \psi_{\pmd}] = \pm \frac{1}{2} \psi_{\pmd}
{}~~~~&,&~~~~[\Mb, \psi_{\pm}]=0  \ .
\ena
The action of the operators on all other quantities can be readily deduced from
these.

Our paper is organized as follows: In Section 2 we present the (2,2) geometry
and the constraints for an extended theory where one gauges {\em both} $U(1)$
tangent space symmetries.
In Section 3  we  obtain their solution. In Section 4 we obtain  the
prepotentials for minimal axial supergravity by restriction to the axial $U(1)$
 tangent space symmetry. Finally, in Section 5 we discuss superWeyl invariance
and
construct the vielbein superdeterminant.

\newpage

\sect{The N=2 geometry}

The geometry of two-dimensional N=2 supergravity has been described by Howe and
Papadopoulos$^{10}$ and  is also discussed by Gates {\em et al}$^{~11}$.
The theory is described by  suitably defined covariant
derivatives which include the action of the tangent space generators introduced
in the previous section.  Suitable constraints on the torsions and curvatures
lead to minimal supergravity multiplets. It is convenient, to begin with,  to
keep the theory fully locally invariant under all the tangent space generators.
The minimal multiplets are obtained, however, by removing either the $U_V(1)$
connection, which leads to what one would obtain directly by dimensional
reduction from four-dimensional supergravity using a chiral compensator, or the
$U_A(1)$ connection, leading to a theory which uses a twisted chiral
compensator. The introduction of an extra $U(1)$ symmetry and its subsequent
degauging is similar to the procedure used in the four-dimensional
situation$^6$.

The spinorial covariant derivatives are defined by
\bea
\Del _{\a} &=& E_{\a}+ \Phi_{\a}\Lambda+\tilde{\Phi}_{\a} \tilde{\Lambda}
+\Sigma_{\a}
N \nonumber\\
&=& E_{\a} +\Omega_{\a}M +\Gamma_{\a}\Mb +\Sigma_{\a}N
\ena
with $\a=\pm$, and corresponding expressions for the complex conjugate
spinorial derivatives as well as the vectorial derivatives.  The vielbein is
given   by
\EQ
E_A= E_A^{~~M} \pa_M  \ .
\EN
Torsions
and curvatures are defined as usual by
\EQ
[\Del _A , \Del_B \} = T_{AB}^{~~~C} \Del_C + R_{AB}M +\Rb_{AB}\Mb
+F_{AB}N  \ .
\EN
They satisfy constraints which can be described by the following
anticommutators$^{10,11}$.
\bea
\{\Del_+ ,\Del_+\} &=&0 ~~~~~~~,~~~~  \{\Del_-, \Del_-\}=0 \nonumber\\
\{\Del_{+} , \Del_{\pd} \} &=&i \Del_{\pp} ~~~~,~~~~\{\Del_{-} , \Del_{\md} \}
= i \Del_{\mm} \nonumber\\
&~& \nonumber\\
  \{\Del_+, \Del_-\} &=& -\frac{\Rb}{2}(\Lambda-i \tilde{\Lambda}) =
-\Rb \Mb  \nonumber\\
\{\Del_+, \Del_{\md} \}&=& F(\Lambda -iN)= F(M+\Mb -iN)
\ena
as well as their complex conjugates.
Additional constraints follow from the use of Bianchi identities. Furthermore,
for the minimal supergravities one restricts the gauge group so that either
$F=0$ for the $U_A(1)$ version,
or $R=0$ for the $U_V(1)$ version, by setting either  $\Sigma_{\a}=0$ or
$\tilde{\Phi}_{\a}= \Omega_{\a} - \Gamma_{\a}=0$$^{~10}$.

We solve the constraints  in (2.4) by expressing the covariant derivatives in
terms of two  (pre)potentials, a  real vector superfield $H^m$ and a  (scale
compensator) complex scalar superfield  $S$. This leads to a description of
nonminimal $U_V(1) \times U_A(1)$ supergravity$^{10}$.
In doing so, we implicitly make  certain supersymmetric gauge choices which
remove a large number of irrelevant superfields by means of algebraic
(ghost-nongenerating) gauge transformations. We also use educated guesses based
on four-dimensional experience to partially determine the dependence on the
prepotentials.

We make a useful observation: the minimal axial version of the theory is
obtained by
setting $\Sigma_{\a} = F=0$ and this additional constraint implies that $S$
satisfies a condition which eventually expresses it in terms of a {\em chiral}
compensator $\phi$. Therefore, axial supergravity is described in terms of the
two prepotentials $H^m$ and $\phi$.
However,  the constraints in (2.4) and their solution are invariant under the
interchange $\Del_- \leftrightarrow \Del_{\md}$ together with $\tilde{\Lambda}
\leftrightarrow N$ and $\Rb \leftrightarrow -2F$.  {\em Therefore, the
minimal vector version of  $(2,2)$ supergravity,
with $\tilde{\Phi}_{\a} = R=0$  is obtained directly from the axial vector
version
by the above substitutions, which imply in particular that the chiral scalar
compensator is replaced by a twisted chiral compensator.}

\sect{Solving the N=2 constraints}

\subsection{Determining the connections}

We begin by defining the ``hat''  differential operators
\EQ
\hat{E}_{\pm}= e^{-H}D_{\pm}e^{H}~~~~,~~~~ H=H^mi\pa_m  \ .
\EN
We note the explicit forms $\Eh _{\pm} = D_{\pm} +iH_{\pm}^m\pa_m$,  where
$H_{\pm}^{m}$ is a function of $H^m$ and its derivatives. Additional features
are $\{\Eh_+,\Eh_+\}=\{\Eh_+,\Eh_-\}=0$ while $\{\Eh_+, \Eh_{\pd} \}\equiv
i\Eh_{\pp} =i\pa_{\pp}+\cdots$~. Thus  $\Eh_{\a}$, $\Eh_{\dot{\a}}$ and
$\Eh_{a}$ form a linearly independent basis
of derivative operators.

In the corresponding four-dimensional case one can find Lorentz gauges in which
the actual spinorial vielbein is proportional to the corresponding ``hat''
object. In two dimensions however the Lorentz group is more restricted and we
postulate instead$^{6,8}$
\EQ
E_+ \equiv e^{\Sb}(\Eh _+ + \Apm \Eh_-) ~~~~,~~~~E_- \equiv  e^{\Sb}(\Eh_-+\Amp
\Eh_+)
\EN
with corresponding expressions for the complex conjugate spinorial vielbein.
Here $\Sb$ is, for the time being, an arbitrary scalar ({\em not} chiral)
superfield, and the $A$'s will be determined later as functions of $H^m$.

We begin by imposing the first constraint, $\{\Del_+, \Del_+\}=0$, which leads
to the conditions
\bea
\{E_+,E_+\} +(\Omega_+ -i\Sigma_+)E_+ &=&0 \nonumber\\
2E_+\Gamma_+ +(\Omega _+- i\Sigma_+)\Gamma_+&=&0 \nonumber\\
2E_+\Omega_+ -i\Sigma_+\Omega_+ &=&0 \nonumber\\
2E_+\Sigma_++\Omega_+\Sigma_+&=&0  \ .
\ena
Substituting in the first equation the explicit expressions for the vielbein we
find, since the $\Eh$
are linearly independent,
\EQ
\Omega_+-i\Sigma_+ =-2 e^{\Sb}(\Eh_+\Sb +\Apm \Eh_-\Sb )
\EN
while the coefficient $\Apm$ must satisfy
\EQ
\Eh_+\Apm +\Apm\Eh_-\Apm =0  \ .
\EN
We will show later on that  our solution for $\Apm$ satisfies this equation.
Notice that the
equation also implies that $\Apm$ is ``linear'' in the sense that
\EQ
\Eh_+\Eh_-\Apm=0 \ .
\EN
The remaining conditions in (3.3) can also be verified once we have
explicit expressions for the connections.

Similar results are obtained from the $\{\Del_-,\Del_-\}=0$ constraint. In
particular we have
\EQ
\Omega_-+i\Sigma_- =2 e^{\Sb}(\Eh_-\Sb+\Amp\Eh_+\Sb )
\EN
and we must satisfy
\EQ
\Eh_- \Amp +\Amp \Eh_+ \Amp =0  \ .
\EN

We turn next to the second constraint, $\{\Del_+,\Del_-\}= -\Rb\Mb$
which leads to
\bea
\{E_+,E_-\}+\frac{1}{2}(\Omega_--i\Sigma_-)E_+-\frac{1}{2}
(\Omega_++i\Sigma_+)E_- &=&0 \nonumber\\
E_+\Omega_- +E_-\Omega_+ -\Omega_+\Omega_-
-\frac{i}{2}\Sigma_+\Omega_--\frac{i}{2}\Sigma_-\Omega_+&=&0 \nonumber\\
E_+\Sigma_-+E_-\Sigma_+
-\frac{1}{2}\Omega_+\Sigma_-+\frac{1}{2}\Omega_-\Sigma_+&=&0
\ena
as well as
\EQ
E_+\Gamma_- +E_-\Gamma_+ -\frac{1}{2}(\Omega_++i\Sigma_+)\Gamma_-
+\frac{1}{2}(\Omega_--i\Sigma_-)\Gamma_+ \equiv -\Rb  \ .
\EN
We substitute in the first equation of (3.9) the explicit expressions for the
vielbein and connections and after some algebra, making use also of  (3.5)  and
(3.8), we find
\bea
\Omega_+&=&+ e^{\Sb}(\Eh_-\Apm - \Apm\Eh_+\Amp )\nonumber\\
\Omega_- &=&-e^{\Sb}( \Eh_+\Amp -\Amp \Eh_-\Apm)  \ .
\ena
The other conditions can be verified once we have explicit expressions for
$\Sigma_{\pm}$.

Finally we turn to the anticommutator
\bea
\{\Del_+,\Del_{\md} \}  &=&
\{E_+,E_{\md}\}-\frac{1}{2}(\Gamma_+-i\Sigma_+)E_{\md}
+\frac{1}{2}(\Gamma_{\md}-i\Sigma_{\md})E_+ \nonumber\\
&+&[E_+\Gamma_{\md}+E_{\md}\Omega_+-\frac{1}{2}\Gamma_+\Gamma_{\md}
+\frac{1}{2}\Gamma_{\md}\Omega_++\frac{i}{2}\Sigma_+\Gamma_{\md}
-\frac{i}{2}\Sigma_{\md}\Omega_+]M \nonumber\\
&+&[E_+\Omega_{\md}
+E_{\md}\Gamma_+-\frac{1}{2}\Gamma_+\Omega_{\md}+
\frac{1}{2}\Gamma_{\md}\Gamma_+ +\frac{i}{2}\Sigma_+
\Omega_{\md}-\frac{i}{2}\Sigma_{\md}\Gamma_+]\Mb \nonumber\\
&+&[E_+\Sigma_{\md}+E_{\md}\Sigma_+-\frac{1}{2}\Gamma_+\Sigma_{\md}+
\frac{1}{2}\Gamma_{\md}\Sigma_++i\Sigma_+\Sigma_{\md}]N \ .
\ena
We require that the first line vanish. Matching terms, and also using the fact
that the anticommutator of  $\Eh$'s does not produce spinorial derivatives, we
find the results
\bea
\Gamma_+-i\Sigma_+&=&+2e^{\Sb}(\Eh_+S+\Apm\Eh_-S) \nonumber\\
\Gamma_{\md}-i\Sigma_{\md}&=&-2e^{S}(\Eh_{\md}\Sb+A_{\md}^{~\pd}\Eh_{\pd}\Sb)
\ena
as well as the conditions
\bea
\Eh_+A_{\md}^{~\pd}+\Apm \Eh_-A_{\md}^{~\pd} &=&0 \nonumber\\
\Eh_{\md}  \Apm +A_{\md}^{~\pd} \Eh_{\pd}\Apm &=&0
\ena
and
\EQ
\{\Eh_+,\Eh_{\md}\}+A_{\md}^{~\pd}\{\Eh_+,\Eh_{\pd}\}+\Apm\{\Eh_-,\Eh_{\md}\}
+\Apm A_{\md}^{~\pd} \{\Eh_-,\Eh_{\pd}\}=0  \ .
\EN
This last condition determines the $A$'s.

At this point all the connections are determined and, provided the conditions
on the $A$'s hold, it is possible to check that all the other conditions we
have encountered are indeed satisfied. In particular one can check that the
second and third lines in (3.12) equal $i$ times the fourth line so that the
last constraint in (2.4) is  obeyed, with
\EQ
F=i[E_+\Sigma_{\md}+E_{\md}\Sigma_+-\frac{1}{2}\Gamma_+\Sigma_{\md}+
\frac{1}{2}\Gamma_{\md}\Sigma_++i\Sigma_+\Sigma_{\md}] \ .
\EN

Besides the $\Omega_{\pm}$ in (3.11) we have, from (3.4) and (3.13),
\bea
\Sigma_+&=&-2ie^{\Sb}(\Eh_+\Sb +\Apm \Eh_-\Sb )-ie^{\Sb}(\Eh_-\Apm -\Apm
\Eh_+\Amp ) \\
\Sigma_-&=&-2ie^{\Sb}(\Eh_-\Sb +\Amp \Eh_+\Sb )-ie^{\Sb}(\Eh_+\Amp -\Amp
\Eh_-\Apm ) \nonumber\\
\Gamma_+ &=& +2e^{\Sb}(\Eh_+S+\Apm \Eh_-S)+2e^{\Sb}(\Eh_+\Sb +\Apm \Eh_-\Sb
)+e^{\Sb}(\Eh_-\Apm -\Apm \Eh_+\Amp ) \nonumber\\
\Gamma_- &=& -2e^{\Sb}(\Eh_-S+\Amp \Eh_+S)-2e^{\Sb}(\Eh_-\Sb +\Amp \Eh_+\Sb
)-e^{\Sb}(\Eh_+\Amp -\Amp \Eh_-\Apm ) \nonumber
\ena
and corresponding expressions for the complex conjugates.

\subsection{Determining $\Apm$ and $\Amp$}

We have defined
\bea
\Eh_{\pp} &\equiv& -i\{\Eh_+,\Eh_{\pd}\}  = \pa_{\pp}+\cdots \nonumber\\
\Eh_{\mm} &\equiv& -i\{\Eh_-,\Eh_{\md}\}  = \pa_{\mm}+\cdots  \ .
\ena
{}From the form of the hatted objects it is clear then that
\bea
\{\Eh_+, \Eh_{\md}\} &=&
\Ch_{+\md}^{~~~\pp}\Eh_{\pp}+\Ch_{+\md}^{~~~\mm}\Eh_\mm
\nonumber\\
\{\Eh_-, \Eh_{\pd}\} &=&
\Ch_{-\pd}^{~~~\pp}\Eh_{\pp}+\Ch_{-\pd}^{~~~\mm}\Eh_\mm
\ena
without any spinor contributions on the right hand side. These equations define
the hatted anholonomy coefficients as power series in the field $H^m$ and its
derivatives.

With these definitions, (3.15) breaks up into two equations
\bea
\Cpmp+i \Ampd+ \Cmpp\Apm \Ampd &=&0 \nonumber\\
\Cpmm+i\Apm + \Cmpm \Apm \Ampd &=&0  \ .
\ena

In a similar manner, from $\{\Del_-, \Del_{\pd} \}$ one finds
\bea
\Cmpp+i \Amp+ \Cpmp\Amp \Apmd &=&0 \nonumber\\
\Cmpm+i\Apmd + \Cpmm \Amp \Apmd &=&0  \ .
\ena

By eliminating one of the unknowns in (3.20)  one is led to  a quadratic
equation, and therefore the
$A$'s can be found explicitly:
\bea
\Apm &=& \frac{ \Cpmp \Cmpm -\Cpmm\Cmpp+1 - \sqrt{ (\Cpmp \Cmpm
-\Cpmm\Cmpp+1)^2 +4 \Cpmm\Cmpp}}{2i\Cmpp} \nonumber\\
\Ampd &=& \frac{ \Cmpp \Cpmm -\Cmpm\Cpmp+1 - \sqrt{(\Cmpp \Cpmm
-\Cmpm\Cpmp+1)^2 +4\Cmpm\Cpmp}}{2i\Cmpm} \nonumber\\
&~&
\ena
We have chosen the signs of the square roots so that the $A$'s vanish when
$H^m$ and therefore the $\hat{C}$'s vanish.

It is obvious that  the solutions of (3.21) are simply related to those of
(3.20). One finds
\EQ
\Amp = \frac{\Cmpp}{\Cpmm}\Apm ~~~~,~~~~ \Apmd = \frac{\Cmpm}{\Cpmp}\Ampd \ .
\EN
Thus, the coefficients $A$ are determined explicitly as power series in $H^m$
and its derivatives.

We are now in a position to check that (3.5) and  the first equation in (3.14)
are satisfied.  We
proceed as follows: We take the commutator of (3.15) first with $\Eh_+$ and
then with $\Eh_-$. Using Jacobi identities for triple commutators  we rewrite
the results in terms of $\Eh_{\pp}$ and $\Eh_\mm$ and set their coefficients to
zero. We find  the following four equations:
\bea
ix+\Cmpm (\Ampd x+ \Apm y)&=&-\Apm(i \Ch_{+\mm}^{~~~\mm} -i \Ampd \Ch_{-
\pp}^{~~~\mm} )\nonumber\\
iy+\Cmpp (\Ampd x+ \Apm y)&=&-\Apm(i \Ch_{+\mm}^{~~~\pp} -i \Ampd \Ch_{-
\pp}^{~~~\pp} )\nonumber\\
iu+\Cmpm (\Ampd u+ \Apm v)&=&i \Ch_{+\mm}^{~~~\mm} -i \Ampd \Ch_{-
\pp}^{~~~\mm}
\nonumber\\
iv+\Cmpp (\Ampd u+ \Apm v)&=&i\Ch_{+\mm}^{~~~\pp} -i \Ampd \Ch_{- \pp}^{~~~\pp}
\ena
where we have temporarily denoted $x= \Eh_+\Apm$, $y= \Eh_+ \Ampd$,
$u=\Eh_-\Apm$ and $v= \Eh_-\Ampd$.  It is obvious then that the solutions of
the last two equations are proportional to the solutions of the first two, with
proportionality factor $-\Apm$, and this fact
expresses the content of (3.5) and (3.14). Obviously  (3.8) and the second
equation in (3.14)
are checked in a similar manner.

With this, and some additional checks on the equations involving the
connections,  we have completed the solution of the constraints for $N=2$,
 $U_V(1) \times U_A(1)$ supergravity, expressing the covariant derivatives
 in terms of the unconstrained superfields $H^m$ (a real vector superfield) and
$S$ (a complex scalar superfield).  The relevant results are contained in
(3.1,2,22,23) and (3.11,17). Additional restrictions arise when we reduce the
theory by eliminating one of the $U(1)$ connections.

\sect{Degauging}

We obtain a minimal supergravity multiplet by imposing an additional
constraint which
requires that either the curvature $F$ or the curvature $R$ in (2.4)
vanish$^{10}$.
In  the first case this can be achieved by setting the connection
$\Sigma_{\pm}$ to zero, while in the second case the condition becomes
$\Omega_{\pm} =\Gamma_{\pm}$. We discuss axial
supergravity, obtained by eliminating the $U_V(1)$ connection
$\Sigma_{\pm}$, but as explained in Section 2 the results for vector
supergravity follow immediately.

The connection $\Sigma_{\pm}$ has been given explicitly in (3.17), and setting
it to zero implies a differential constraint on the scalar superfield $S$. It
is not immediately obvious how this constraint is to be solved and therefore we
derive a different, and much more suitable, expression for this connection,
following techniques similar to those used in four dimensions as described in
ref.~6 (in particular  the derivation of eq.~(5.3.25), as well
as subsection (5.3.b.4)). However, the manipulations we perform, though
elementary, are rather baroque and the impatient reader may wish to skip
directly to the final result in eq.~(4.21).

It is convenient to define ``checked'' operators $\Ec$ by
\EQ
\Ec_{\pm} \equiv E_{\pm} ~~~~,~~~~ \Ec_{\pp} \equiv -i \{\Ec_+ , \Ec_{\pd}\}
{}~~~~,~~~~ \Ec_\mm \equiv -i\{\Ec_- , \Ec_{\md} \}
\EN
and corresponding ``checked'' anholonomy coefficients. In particular, using
Jacobi identities, from
\bea
\{\Ec_+, \Ec_{\pp}\} &=&  -\frac{1}{2} (\Omega_+-i\Sigma_+)\Ec_{\pp}
-\frac{i}{2}[ E_{\pd}(\Omega_+-i\Sigma_+)] \Ec_ +\nonumber\\
\{\Ec_+, \Ec_{\mm}\} &=&\frac{1}{2}(\Omega_++\Gamma_+ )\Ec_\mm + {\rm spinorial
{}~vielbeins}
\ena
we identify
\EQ
\Omega_+-i\Sigma_+ = -2 \Cc_{+\pp}^{~~~\pp} ~~~~,~~~~ \Omega_+ +\Gamma_+ = 2
\Cc_{+\mm}^{~~~\mm}
\EN
so that
\EQ
\Gamma_++i\Sigma_+= 2(\Cc_{+\pp}^{~~~\pp} +\Cc_{+\mm}^{~~~\mm}) =
2\Cc_{+a}^{~~~a} \ .
\EN
On the other hand, from the vanishing of the first line in  (3.12) it is
evident that
 \EQ
\Gamma_+-i\Sigma_+ = 2\Cc_{+\md}^{~~~\md}
\EN
so that
\EQ
\Sigma_+= i\Cc_{+\md}^{~~~\md} -i \Cc_{+a}^{~~~a}  = -i[ (-1)^B \Cc_{+B}^{~~~B}
+ \Cc_{+\a}^{~~~\a}]
\EN
since obviously, from the definitions in (4.1), $\Cc_{+\pd}^{~~~\pd} =0$. In
the equation above $(-1)^B$ denotes the usual graded sum.

We also note, from $\Del_{\pp}=-i\{\Del_+, \Del_{\pd}\} = E_{\pp} +
connections$, that
\EQ
E_{\pp} = \Ec_{\pp}
-\frac{i}{2}(\Gamma_++i\Sigma_+)E_{\pd}-\frac{i}{2}
(\Gamma_{\pd}-i\Sigma_{\pd})E_+
\EN
from which one can deduce that the vielbein superdeterminant equals the
``checked'' superdeterminant
\EQ
E={\rm sdet}E_A^{~M}= {\rm sdet}\Ec_A^{~M} \equiv \Ec \ .
\EN
Furthermore, we have  ($z^M=(x^m, \th^{\m}, \th^{\dot{\m}})$)
\EQ
(-1)^B \Cc_{AB}^{~~~B} = \Ec_M^{~B}[\Ec_A,\Ec_B\}z^M=
\Ec_M^{~B}\Ec_A\Ec_B^{~M}-\Ec_M^{~B}\Ec_B\Ec_A^{~M}
\EN
where the summation over repeated indices is graded (but has not been
explicitly indicated, for notational simplicity).

Following techniques in subsection (5.3.b.4) of ref.~6 we
define
adjoint operators such as ${\BAE}_A $ or $\BAH$ by, for example,
\EQ
X\BAE_A = X \Ec_A^{~M}{\buildrel \leftarrow \over  \pa} _M = \pa_M
(X\Ec_A^{~M})
\EN
again with appropriate grading.
They obey the Leibnitz rule
\EQ
XY\BAE_A=X[Y,\BAE_A]+X \BAE_AY =X\Ec_AY+X\BAE_AY
\EN
and are extremely useful for the operations that follow.

Thus we rewrite (4.9) as
\bea
(-1)^B\Cc_{AB}^{~~~B}&=&\Ec_M^{~B}\Ec_A\Ec_B^{~M}-\Ec_M^{~B}\Ec_A^{~M}\BAE_B
+\Ec_M^{~B}\BAE_B \Ec_A^{~M}\nonumber\\
&=& \Ec_A \ln \Ec -\bone \cdot {\BAE_A} + 0
\ena
the first term being a standard expression for the derivative of the
determinant and the last term vanishing since $\Ec_M^{~B}
\BAE_B=\Ec_M^{~B}\Ec_B^{~N} {\buildrel \leftarrow \over \pa}_N =\bone
\cdot{\buildrel
\leftarrow \over \pa}_M =0$.
The right-hand-side can be rewritten, using (4.11),  as
\EQ
-E\Ec_A E^{-1} - \bone \cdot\BAE_A= -E^{-1}\BAE_A E
\EN
so that,  from (4.6)
\EQ
\Sigma_+ = iE^{-1}\BAE_+E-i\Cc_{+\a}^{~~~\a} \ .
\EN

Now, from the first equation in (3.3) and (3.9),
\EQ
\Cc_{++}^{~~~+}=-(\Omega_+-i\Sigma_+)
{}~~~~,~~~~\Cc_{+-}^{~~~-}=\frac{1}{2}(\Omega_++i\Sigma_+)
\EN
and this allows us to write, substituting expressions from (3.11) and (3.17),
\bea
\Sigma_+ &=& iE^{-1}\BAE_+E +\frac{i}{2}\Omega_++\frac{3}{2}\Sigma_+ \\
&=&i E^{-1} e^{\Sb}   [\BAEh_++\Apm \BAEh_-]E
 -ie^{\Sb}(\Eh_-\Apm -\Apm\Eh_+\Amp ) -3 iE_+ \Sb\nonumber
\ena
or
\EQ
-i\Sigma_+ =\bone \cdot [ \BAEh_+ + \BAEh_-\Apm]
e^{\Sb}+e^{\Sb}\Apm\Eh_+\Amp - E_+\ln E -2E_+ \Sb
\EN
where the right-hand-side is obtained after a number of intermediate steps
using repeatedly the Leibnitz formula for the operators $\BAEh_{\pm}$.

 We write
\bea
\bone \cdot\BAEh_+ &=&\bone \cdot e^{\BAH} \BAD_+e^{-\BAH}= \bone
\cdot[D_+(\bone \cdot e^{\BAH})]e^{-\BAH}
\nonumber\\
&=& [\bone \cdot e^{-\BAH}]e^{-H}D_+e^{H}[\bone \cdot e ^{-\BAH}]^{-1} =
[\bone \cdot e^{-\BAH}]\Eh_+[\bone \cdot e^{-\BAH}]^{-1}
\nonumber\\
&=& -\Eh_+[ \ln (\bone \cdot e^{-\BAH})]
\ena
where in the second line we have used again the Leibnitz rule (in exponential
form) as well as  the identity $\bone =(\bone \cdot e^{- \buildrel \leftarrow
\over  X})e^{
\buildrel \leftarrow \over  X}=(\bone \cdot e^{ \buildrel \leftarrow \over
X})[e^X(\bone \cdot e^{- \buildrel \leftarrow \over  X})]$ (see ref.~6
eq.~(5.3.51b)).
With a similar expression for $\bone \cdot\BAEh_-$, the first term in (4.17)
becomes
\hbox{$-E_+[\ln (\bone \cdot e ^{-\BAH})]$}.

Using the identity (3.8)  and the explicit expression for $E_+$ we also rewrite
\EQ
e^{\Sb}\Eh_+\Amp= \frac{1}{1-\Apm \Amp} E_+\Amp
\EN
and since, by (3.5),  $E_+\Apm =0$,
\EQ
e^{\Sb}\Apm \Eh_+ \Amp = \frac {E_+(\Apm \Amp)}{1-\Apm \Amp} = -E_+ \ln (1-\Apm
\Amp )  \ .
\EN

Substituting (4.18) and (4.20) into (4.17), and using similar manipulations for
$\Sigma_-$ leads us to the final form for the $U_V(1)$ connections:
\EQ
i\Sigma_{\pm} = E_{\pm} \ln \left[ (\bone \cdot e^{-\BAH})E e^{2\Sb}(1-\Apm
\Amp )
\right]
\EN
with similar expressions for the complex conjugates.

The degauging  of the $U_V(1)$ symmetry takes place by simply requiring the
connections $\Sigma_{\pm}$  and their complex conjugates to vanish.  Setting
the expression in (4.21) to zero implies that the quantity in the square
bracket is a (covariantly) antichiral scalar superfield. We write it  as $e^{2
\bar{\s}}$ where the superfield $\bar{\s}$ satisfies the condition
\EQ
\Eh_{+}\bar{\s}= \Eh_{-}\bar{\s}=0
\EN
i.e.
\EQ
\bar{\s} = e^{-H}\bar{\phi} ~~~~,~~~~ D_{\pm}\bar{\phi} =0
\EN
so that  ${\phi}$ is an ordinary chiral field.
Solving for the compensator $\Sb$ we have then
\EQ
e^{\Sb} = e^{\bar{\s}} \frac{ \left[\bone \cdot e^{-\BAH}
 \right]^{-\frac{1}{2}}}{[1-\Apm \Amp]^{\frac{1}{2}}} E^{-\frac{1}{2}}
\EN
and substituting back into the vielbein gives the solution  of the constraints
for minimal axial $(2,2)$ supergravity in terms of the
superfields $H^m$ and $\s$ (or $\phi$).

In view of the remarks at the end of Section 2, the solution for the vector
version is immediate.
One performs the interchange $- \leftrightarrow \md$ everywhere, starting
with the definitions in (3.1). One has exactly the same form for the solution,
except that the compensator satisfies
\EQ
\Eh_+ {\bar\s} =\Eh_{\md}\bar{\s} = 0
\EN
so that the corresponding $\bar{\phi}$  satisfies $D_+ \bar{\phi} =
D_{\md}\bar{\phi}=0$, i.e. $\phi$ is an ordinary twisted chiral superfield.

This completes the solution of the constraints.

\sect{SuperWeyl scaling and the vielbein superdeterminant}

As is well-known, the constraints in (2.4) have an additional invariance under
scaling
with an {\em arbitrary} scalar superfield $L$ (which  can be restricted to be
real since imaginary scale transformations can be absorbed into the $U_V(1)$
transformations). The scaling can be implemented as a shift in the scale
compensator, $S \rightarrow S+L$. In particular we have then the following
transformation properties:
\bea
E_{\pm} &\rightarrow& e^L E_{\pm} \nonumber\\
\Omega_{\pm} &\rightarrow& e^L \Omega_{\pm} \nonumber\\
\Sigma _{\pm}&\rightarrow&e^L\left( \Sigma_{\pm} -2iE_{\pm}L\right) \nonumber\\
\Gamma_{\pm} &\rightarrow& e^L\left(\Gamma_{\pm} \pm 4E_{\pm}L\right)
\ena
and it is possible  to check explicitly the invariance of the constraints with,
in particular
\bea
\Rb &\rightarrow& e^{2L}(\Rb +4[\Del_-,\Del_+]L) \nonumber\\
F &\rightarrow& e^{2L}(F-2i[\Del_{\md},\Del_+]L)
\ena
However, after degauging, the scale parameter must be restricted.

We note that under  combined infinitesimal scale and $U_V(1)$ gauge
transformation with
parameter $\n$  (i.e. $\delta \Del_{\pm}  =[ \n N, \Del_{\pm}]$), the
connection $\Sigma_{\pm}$, in addition to the overall rescaling,
shifts:
\EQ
\Sigma_{\pm} \rightarrow  (1+L-\frac{i}{2}\nu )\Sigma_{\pm} -2i
E_{\pm}(L-\frac{i}{2}\n )
\EN
Therefore, after degauging,  the axial version of the theory will only be
invariant under (combined) transformations with parameter $\bar{\lambda} =
L-\frac{i}{2}\n$ which maintain the
vanishing of $\Sigma_{\pm}$, i.e. complex scale transformations with
covariantly (anti)chiral parameter $\bar{\l}$
satisfying
\EQ
E_{\pm}\bar{\l} =0
\EN
It is easy to see, from  (4.24), that these are implemented by the shift of the
chiral compensator,
\EQ
\s \rightarrow \s +\l \ .
\EN
Obviously, in the vector version of the theory the corresponding superWeyl
parameter is twisted chiral$^{10}$.

To complete this work, we compute the superdeterminant $E = {\rm sdet}
E_A^{~M}$, where
$E_A= E_A^{~M} \pa_M$. As already stated in (4.8)  we have $E = \check{E}$. It
is convenient to write
\EQ
\check{E_A}= \check{E}_A^{~B} \Eh_B  ~~~~,~~~~
\Eh_B= \Eh_B^{~M} \pa_M
\EN
so that
\EQ
E= \check{E}= {\rm sdet} \check{E}_A^{~B} \cdot {\rm sdet} \Eh_B^{~M} \ .
\EN
Writing
\bea
\Eh_{\pm} &=& e^{-H}D_{\pm}e^{H} = D_{\pm} +iH_{\pm}^m \pa_m \nonumber\\
\Eh_{\pmd} &=&e^{H}D_{\pmd}e^{-H} = D_{\pmd}-iH_{\pmd}^m \pa_m \ ,
\ena
and also
\bea
\Eh_{\pp}&=  & -i\{\Eh_+ , \Eh_{\pd} \}= \pa_{\pp} +[-D_+H_{\pd}^n
+D_{\pd}H_+^n -iH_+^m \pa_mH_{\pd}^n-iH_{\pd}^m\pa_mH_+^n]\pa_n \nonumber\\
\Eh_{\mm}  &=&-i\{\Eh_- , \Eh_{\md}\}= \pa_{\mm} +[-D_-H_{\md}^n +D_{\md}H_-^n
-iH_-^m \pa_mH_{\md}^n-iH_{\md}^m\pa_mH_-^n]\pa_n  \nonumber\\
&~&
\ena
we note that $\Eh_B^{~M}$ is block triangular and therefore
$$\displaylines{
\quad -{\rm sdet} \Eh_B^{~M} = \hfill\stepcounter{equation}(\theequation)\cr
\left |\matrix{
1-D_+H_{\pd}^{\pp}+
D_{\pd}H_+^{\pp}-iH_+^m\pa_mH_{\pd}^{\pp}-iH_{\pd}^m\pa_mH_+^{\pp}
&
-D_+H_{\pd}^\mm +D_{\pd}H_{\pd}^\mm -iH_+^m\pa_mH_{\pd}^\mm
-iH_{\pd}^m\pa_mH_+^\mm
\cr
-D_-H_{\md}^{\pp}+D_{\md}H_-^{\pp}
-iH_-^m\pa_mH_{\md}^{\pp}+H_{\md}^m\pa_mH_-^{\pp}
&
1-D_-H_{\md}^\mm+D_{\md}H_-^\mm
-iH_-^m\pa_mH_{\md}^{\mm}-iH_{\md}^m\pa_mH_-^{\mm}
\cr
}\right | \cr
}$$
which can be computed to any order in powers of $H^m$ and its derivatives.

Turning to the computation of ${\rm sdet} \check{E}_A^{~B}$, with $E_{\pm}$,
$E_{\pmd}$ given in (3.2) and $\Ec_{\pp}$, $\Ec_\mm$ defined in (4.1), we
note again that $\Ec_A^{~B}$ is block triangular so that in $\Ec_{\pp}$,
$\Ec_\mm$ we only need the pieces proportional
to $\Eh_{\pp}$,  $\Eh_\mm$, e.g.
\bea
\Ec_{\pp} &=& -i\{E_+ , E_{\pd}\} = \{ e^{\Sb}(\Eh_+ +\Apm \Eh_-), e^S
(\Eh_{\pd}+ \Apmd \Eh_{\md}) \} \\
&=&-i e^{(S+\Sb)} \left[ \{\Eh_+,\Eh_{\pd}\} +\Apm \{\Eh_-,\Eh_{\pd} \} +\Apmd
\{\Eh_+,\Eh_{\md}\}
+\Apm \Apmd\{\Eh_-, \Eh_{\md} \} + \cdots \right] \nonumber\\
&=& -ie^{(S+\Sb)}\left[ (i+\Apm \Cmpp +\Apmd\Cpmp )\Eh_{\pp} +(i\Apm\Apmd +\Apm
\Cmpm +\Apmd \Cpmm )\Eh_\mm +\cdots \right] \nonumber
\ena
and the superdeterminant works out to be
\bea
&&{\rm sdet} \Ec_A^{~B} =
- \frac{
\left |\matrix{
i+\Apm \Cmpp +\Apmd \Cpmp
 &
i\Apm \Apmd+\Apm\Cmpm +\Apmd\Cpmm
\cr
i\Amp \Ampd +\Amp\Cpmp+\Ampd\Cmpp
 &
i+\Amp\Cpmm +\Ampd\Cmpm
 \cr
}\right | \downprop}{[1-\Apm\Amp ][ 1-\Apmd \Ampd ]} \nonumber
\ena
The numerator can be simplified by using (3.20,21,23). After some algebra we
obtain as a final form
$$\displaylines{
\quad {\rm sdet} \Ec_A^{~B}= \hfill\stepcounter{equation}(\theequation)\cr
\left[(1+\Cpmp\Cmpm +\Cpmm\Cmpp )^2-4\Cpmp \Cmpm \Cpmm \Cmpp \right]\frac{
1-\Apm \Amp \Apmd \Ampd}{(1-\Apm\Amp )(1-\Apmd \Ampd)} \cr
}$$
and again this can be computed to any order in powers of $H^m$ and its
derivatives.
Note that, as expected, these superdeterminants have no dependence on the scale
compensators, reflecting the classical scale invariance of the theory
described by the action $\int d^2x d^4 \th E^{-1}$.

\sect{Conclusions}

We have presented in this work the solution to the constraints of
two-dimensional, $(2,2)$ supergravity, expressing the constrained covariant
derivatives
(vielbein and connections) in terms of two  unconstrained prepotentials, $H^m$
and $\phi$,
where $H^m$ is a real vector superfield and $\phi$ is an ordinary chiral,
or twisted chiral, scalar superfield. Our results are contained in eqs.
(3.1,2,22,23) and (4.24), and (3.11,17). The results are not as simple as
they are in  $(1,0)$$^3$ or $(1,1)$$^4$
supergravity, but we hope they will be useful  for some studies where a fully
covariant $N=2$ formalism may prove advantageous. These might include on one
hand  topics concerning $N=2$ superRiemann surfaces, and on the
other issues dealing with the quantization of induced (2,2) supergravity and
its interaction with (2,2) matter. Further topics, such as chiral
representation, behavior under
supercoordinate
transformations, the background field method and quantization issues will be
presented
elsewhere$^9$.

\def\thesubsection{} 
\subsection{Acknowledgements}

Marcia Wehlau thanks the Physics Department of Queen's University for
its hospitality.


\end{document}